\documentclass[a4paper,12pt]{article}
\usepackage[top=25truemm,bottom=25truemm,left=15truemm,right=15truemm]{geometry}
\usepackage[]{hyperref}
\usepackage{amsmath}
\usepackage{ascmac}
\usepackage{amssymb}
\usepackage[]{graphicx,caption}
\graphicspath{{./figs/}}

\title{Interface in AdS black hole spacetime}
\author{Koichi Nagasaki}
\date{\today}

\begin{document}
\vspace{1cm}
\begin{center}
{\LARGE Interface in AdS black hole spacetime}\\
\vspace{2cm}
{\large Koichi Nagasaki}\footnote{koichi.nagasaki24@gmail.com}\\
\vspace{1cm}
{\small School of Physics,
University of Electronic Science and Technology of China,\\
Address: No.4, Section 2, North Jianshe Road, Chengdu 610054, China}
\end{center}
\vspace{1.5cm}
\abstract{
We consider a defect solution in the AdS black hole spacetime.
This is a generalization of the previous work \cite{Nagasaki:2011ue} to another spacetime.
This system consists of D3 and D5 branes.
The equation of motion for a sort of non-local operator, ``an interface," is given and its numerical solution is shown by the numerical calculation.

We also consider a string extends between this interface and the boundary of the AdS spacetime.
This corresponds the quark-interface potential in the boundary theory of the bulk black hole spacetime.
This result gives a new example of holographic relation which includes the gauge flux in the probe D5-brane.
}
\vspace{1cm}

\tableofcontents

\section{Introduction}
In \cite{Nagasaki:2011ue} we considered a sort of non-local operator called ``an interface."
This system consists of the multiple D3-branes and a probe D5-brane.
Here the D3-branes form the AdS$_5\times S^5$ spacetime.
In this previous work we solved the equation of motion of the probe D5-brane and obtained its solution. 
This corresponds to the fuzzy funnel in the gauge theory side.
The result showed that these nonlocal operators introduce a parameter which enables us to compare the results from gauge theory and gravity theory.
This new parameter is related to the gauge flux on the D5-brane.

In this paper we would like to generalize the background spacetime to black hole spacetime.
The result of this paper will be useful for future works about the AdS/CFT correspondence.
For our system including the interface, the whole action consists of the Einstein-Hilbert term and the D5-brane term.
In order to find the effect of the probe brane we focus on the latter term in this paper.
From this action we obtain the equation of motion of the D5-brane and obtain its solution.
This reveals the black hole mass dependence of the form of the interface and also the flux dependence. 

A new point of this paper is the introduction of the flux on the probe brane system.
The gauge flux is used also in the context of the compactification \cite{Koerber:2007jb}.
An importance of the gauge flux becomes clear in comparing holographic quantities.
In this paper \cite{Nagasaki:2011ue} we showed such a comparison in the context of the D3/D5 system. 
It consists of multiple D3-branes and a probe D5-brane.
In the boundary theory it describes a gauge theory with a defect called the interface.
In general the gauge theory calculation is simpler in small 't Hooft coupling $\lambda$ region while the gravitational calculation is simple in large coupling region.
A new parameter relating gauge flux, $k$, enables us to compare the gauge theoretical and gravitational results.
The holographic results are obtained in the power series of $\lambda/k^2$.
By tuning the value of $k$ we can suppress the value $\lambda/k^2$ although $\lambda$ is large.
We can compare the results in the leading order of power series of $\lambda/k^2$.

We also calculate the action of the fundamental string ending on the D5-brane.
This corresponds the potential energy between a quark and the interface on the boundary gauge theory.
It gives the confinement energy \cite{Witten:1998zw, Semenoff:2018ffq, Patrascu:2018ygw, Hanada:2018zxn}.

I would also like to mention about related topics.
This is the motivation to consider the black hole spacetime.
In recent physics Complexity-Action (CA) relation \cite{Brown:2015bva,Brown:2015lvg} is interest and important theme. 
An important concept to study inside the horizon is ``complexity''  \cite{Susskind:2014moa}.
Complexity is very important for studying the information paradox or firewall problem.
According to CA relation, complexity is equal to the action calculated in the region called the Wheeler-DeWitt (WDW) patch.
This is a certain spacetime region defined by a given time.

This holographic relation is studied also in the gauge theories.
The definition of this quantity from gauge theory side is unknown and studied actively, for example, in fermionic theory \cite{Khan:2018rzm}.
To this end the properties of complexity are demanded \cite{Gan:2017qkz, Jefferson:2017sdb, Chapman:2018dem, Chapman:2018hou, Guo:2018kzl, Chapman:2018lsv}.
We studied the effect of the fundamental string for the complexity growth in the previous works \cite{Nagasaki:2017kqe, Nagasaki:2018csh}.
The study of complexity by using nonlocal operators is very useful method to reveal unknown properties of complexity because complexity is known to be non-local \cite{Ageev:2014nva, Fu:2018kcp}.
The interface we study in this paper is a kind of nonlocal operator which has one codimension.
This operator divides the whole space into two gauge theories which have two different gauge groups and studied recently, for example, in \cite{Gutperle:2018fkz, Ovgun:2018jbm}.
In our past work \cite{Nagasaki:2011ue} we studied the interface solution in spacetime AdS$_5\times S^5$. 

This paper is organized as follows.
In section \ref{sec:eom_iAdSSchBH} we set up the AdS$_5\times S^5$ black hole spacetime and find the equation of motion of the interface.
Following the same method used in \cite{Nagasaki:2011ue} we obtain the equation which the interface obeys.
We show the solution of this equation by operating the numerical calculation in section \ref{sec:iAdSBH_sol}.
Following them we calculate the D5-brane action of this interface in section \ref{sec:DBIact}.
This action is defined by integrating on the specific region called the WDW patch.
This calculation is very important for studying the effect to complexity.
Furthermore we calculate the action for the fundamental string ending on the interface in section \ref{sec:iAdSBH_fstring}.
We treat the case where there is no flux because we find this is only the case the probe D5-brane enter the black hole horizon.

\section{Action and Equation of motion}\label{sec:eom_iAdSSchBH}
Here we would like to find the interface solution in black hole spacetime. 
In supergravity side this interface corresponds to the probe D5-brane.
In this section we obtain the equation of motion of this probe D5-brane following the method in \cite{Nagasaki:2011ue}.

The action consists of two terms.
These are the Dirac-Born-Infeld term and the Wess-Zumino term:
\begin{equation}\label{eq:AdSBH_S_D5}
S_\text{D5} = -T_5\int\sqrt{-\det(G+\mathcal F)} + T_5\int\mathcal F\wedge C_4.
\end{equation}
The metric is 
\begin{align}\label{eq:dsAdSBH}
ds_\text{}^2
&= -f(r)dt^2 + \frac{dr^2}{f(r)} + r^2d\Omega_3^2 + d\Omega_5^2,\\
f(r) &= 1 - \frac{r_\text{m}^2}{r^2} + r^2,\;
r_\text{m}^2 = \frac{8M}{3\pi},\nonumber
\end{align}
where $d\Omega_5^2$ is the metric on $S^5$ and $d\Omega_3^2$ is another sphere which is a subspace of AdS$_5$.
The worldvolume of the D5-brane is spanned by coordinates,
\begin{equation}
t,r,\vartheta,\psi_1,\psi_2,\theta,\phi;\qquad
\vartheta,\; \psi_1 \in[0,\pi],
\psi_2\in[0,2\pi),
\end{equation}
where $t$, $r$, $\vartheta$, $\psi_1$ and $\psi_2$ are the coordinates on $AdS_5$ spacetime and $\theta$ and $\phi$ are the coordinates on $S^2$ subspace of $S^5$.
In these coordinates the $S^3$ metric is explicitly
\begin{equation}
d\Omega_3^2 = d\vartheta^2 + \sin^2\vartheta d\psi_1^2 
	+ \sin^2\vartheta\sin^2\psi_1 d\psi_2^2.
\end{equation}
We assume that $\vartheta$ is a function of $r$,
\begin{equation}\label{eq:assump_interface}
\vartheta = \vartheta(r),
\end{equation}
and there is the gauge flux
\begin{equation}
\mathcal F = -\kappa d\theta\wedge(\sin\theta d\phi).
\end{equation}
The Wess-Zumino term consists of this gauge flux and the Ramond-Ramond 4-from, 
\begin{equation}
C_4 
= - r^4dt\wedge d\vartheta\wedge(\sin\vartheta d\psi_1)\wedge(\sin\vartheta\sin\psi_1 d\psi_2) + 4\alpha_4,
\end{equation}
where $\alpha_4$ is a 4-form which satisfies $d\alpha_4=$ (volume of $S^5$).
From these assumptions the induced metric is 
\begin{align}
ds_\text{ind}^2
&= -f(r)dt^2 + \Big(\frac{1}{f(r)}+r^2\vartheta'(r)^2\Big)dr^2 
  + r^2\sin^2\vartheta d\Omega_2^2
  + d\theta^2 + \sin^2\theta d\phi^2,\\
d\Omega_2^2 
&:= d\psi_1^2 + \sin^2\psi_1^2d\psi_2^2.
\end{align}
Adding the gauge flux, its determinant is 
\begin{align}
G+\mathcal F 
&= 
\begin{bmatrix}
-f(r) & & & \\
 & 1/f(r) + r^2\vartheta'^2& & & &\\
 & & r^2\sin^2\vartheta& & &\\
 & & & r^2\sin^2\vartheta\sin^2\psi_1& &\\
 & & & & 1& -\kappa\sin\theta\\
 & & & & \kappa\sin\theta& \sin^2\theta
\end{bmatrix},\nonumber\\
&\Rightarrow
\sqrt{-\det[G+\mathcal F]}
= r^2\sin\theta(\sin^2\vartheta\sin\psi_1)\sqrt{(1+\kappa^2)(1+r^2\vartheta'^2f(r))}
= -\mathcal L_\text{DBI}/T_5.
\end{align}
The Wess-Zumino term is from the gauge flux and the Ramond-Ramond 4-form
\begin{equation}
S_\text{WZ}/T_5
= 4\pi T\kappa\int d\vartheta r^4V(\vartheta)
= (4\pi)^2T\kappa\int dr(\sin^2\vartheta)r^4\vartheta'(r),
\end{equation}
where $T$ is the integration along the time direction and $V(\vartheta)$ is the volume of $S^2$ with radius $\sin\vartheta$ which is $\sin^2\vartheta$ times $4\pi$.

We change the variables as $z = 1/r$ and $\varphi = \pi/2 - \vartheta$.
Introducing a parameter expression $(r(\tau),\varphi(\tau))$, the Lagrangian is
\begin{align}
\frac{S_\text{D5}}{T_5(4\pi)^2T} 
&= -\int d\tau\mathcal L,\;\;
\mathcal L = \frac{\sqrt{1+\kappa^2}}2\frac{\cos(2\varphi)}{z^4}
(\sqrt{\dot z^2+g(z)\dot\varphi^2} - K\dot\varphi),\\
g(z) &:= z^2f(z) = 1+z^2-r_\text{m}^2z^4,\\
K &:= \frac{\kappa}{\sqrt{1+\kappa^2}},
\end{align}
where for convenience we defined $K$ which represents the strength of the gauge flux.
It takes values from 0 to 1.
 
The equations of motion are
\begin{subequations}
\begin{align}
z:&\;\;
\frac{d}{d\tau}\Big(\frac{\cos^2\varphi}{z^4}\frac{\dot z}{\sqrt{\dot z^2+g\dot\varphi^2}}\Big)
+ 4\frac{\cos^2\varphi}{z^5}(\sqrt{\dot z^2+g\dot\varphi^2} - K\dot\varphi)
- \frac{\cos^2\varphi}{z^4}\frac{g'(z)\dot\varphi^2}{2\sqrt{\dot z^2+g\dot\varphi^2}} = 0,\\
\varphi:&\;\;
\frac{d}{d\tau}\Big(\frac{\cos^2\varphi}{z^4}\Big(\frac{g\dot\varphi}{\sqrt{\dot z^2+g\dot\varphi^2}} - K\Big)\Big)
+ \frac{2\sin\varphi\cos\varphi}{z^4}(\sqrt{\dot z^2+g\dot\varphi^2} - K\dot\varphi) = 0.
\end{align}
\end{subequations}
Define $u_z := dz/d\tau$, $u_\varphi := d\varphi/d\tau$ and use the constraint $u_z^2+gu_\varphi^2 = 1$ to fix gauge degrees of freedom.
The equations are
\begin{subequations}\label{eq:iAdSBH_EOM}
\begin{align}
\frac{dz}{d\tau}
&= u_z,\\
\frac{d\varphi}{d\tau}
&= u_\varphi,\\
\frac{du_z}{d\tau}
&= 2u_zu_\varphi\tan\varphi
- \frac{4u_\varphi}{z}(gu_\varphi - K)
+ \frac12g'u_\varphi^2,\\
\frac{du_\varphi}{d\tau}
&= - \frac{2u_z^2}{g}\tan\varphi
 + \frac{4u_z}{zg}(gu_\varphi - K)  -\frac{g'}{g}u_zu_\varphi.
\end{align}
\end{subequations}
We can check the validity of the constraint, $u_z^2+gu_\varphi^2 = 1$, as follows:
By multiplying $1/u_\varphi$ to the third equation and $g/u_z$ to the forth equation, 
\begin{equation*}
\frac1{u_\varphi}\frac{du_z}{d\tau}
= 2u_z\tan\varphi
- \frac{4}{z}(gu_\varphi - K)
+ \frac12g'u_\varphi,\;\;
\frac{g}{u_z}\frac{du_\varphi}{d\tau}
= - 2u_z\tan\varphi
 + \frac{4}{z}(gu_\varphi - K) - g'u_\varphi.
\end{equation*}
By adding them, 
\begin{equation*} 
\frac1{u_\varphi}\frac{du_z}{d\tau}
+ \frac{g}{u_z}\frac{du_\varphi}{d\tau}
+ \frac{g'}2u_\varphi = 0\qquad
\Rightarrow\qquad
u_z\frac{du_z}{d\tau}
+ gu_\varphi\frac{du_\varphi}{d\tau}
+ \frac{g'}2u_zu_\varphi^2 
= \frac12\frac{d}{d\tau}(u_z^2 + gu_\varphi^2) = 0.
\end{equation*}
Then $u_z^2 + gu_\varphi^2$ is constant as we defined it as unity.

\section{Interface solution}\label{sec:iAdSBH_sol}
The solution of differential equations \eqref{eq:iAdSBH_EOM} is obtained by the numerical calculation. 
It gives the embedding of the probe D5-brane in AdS$_5\times S^5$ spacetime.
For the calculation we impose the boundary condition so that the D5-brane behaves in the same way to the solution in Ref. \cite{Nagasaki:2011ue}, $\varphi = \kappa z$, because far from the black hole its gravitational effect can be negligible.

Near the boundary, $r\rightarrow\infty\; (z\approx0)$, the metric \eqref{eq:dsAdSBH} has the singularity. 
In order to avoid it, we choose a small value $z=z_0\approx 0$ and begin the numerical calculation with the initial condition $\varphi_0 = \kappa z_0$.
 
\begin{figure}[h]
	\begin{minipage}[t]{0.5\linewidth}
	\includegraphics[width=\linewidth]{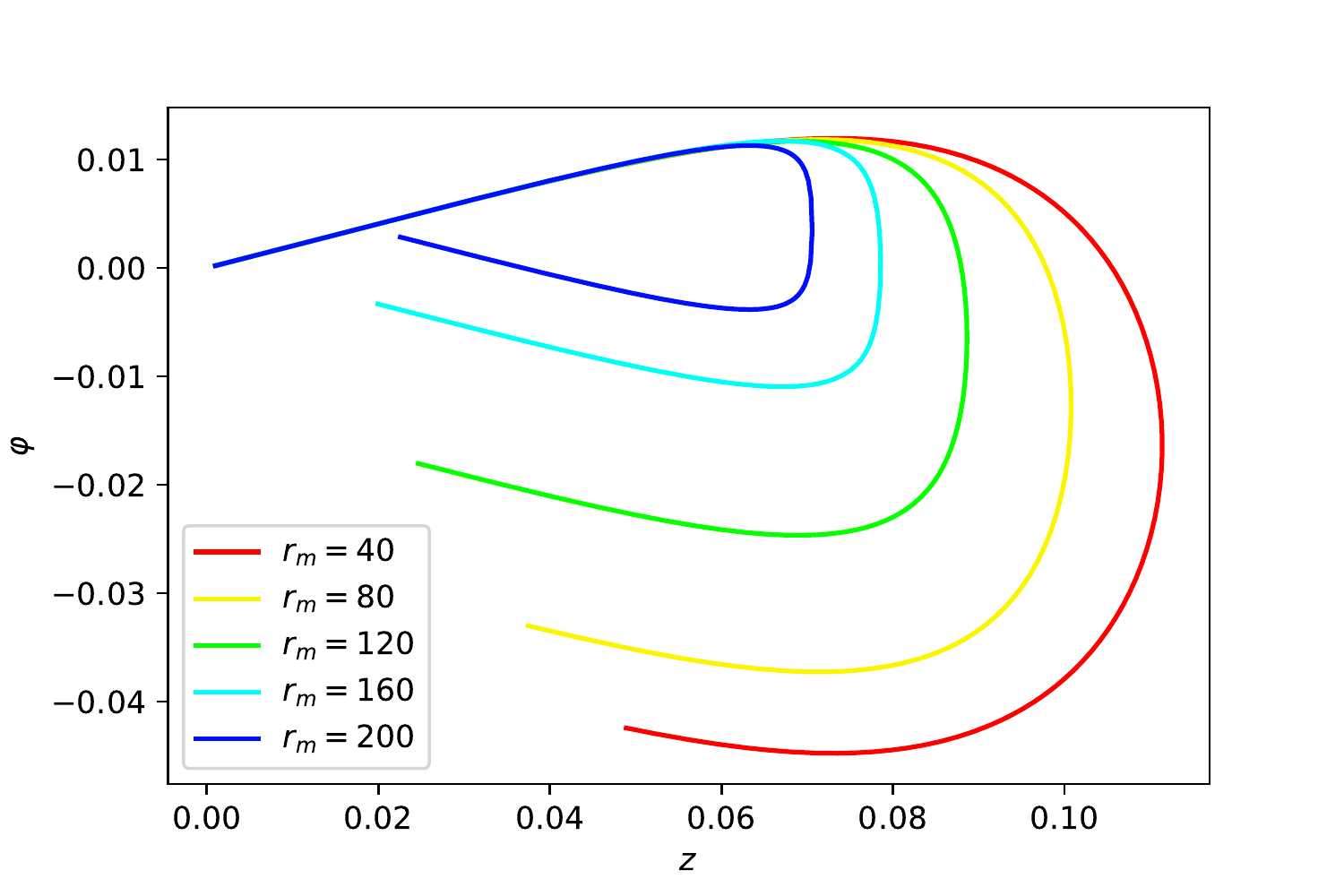}
	\caption{Mass dependence ($K = 0.2$)}
	\label{fig:iAdSBH_hyp_M1}
	\end{minipage}
\hspace{0.01\linewidth}
	\begin{minipage}[t]{0.5\linewidth}
	\includegraphics[width=\linewidth]{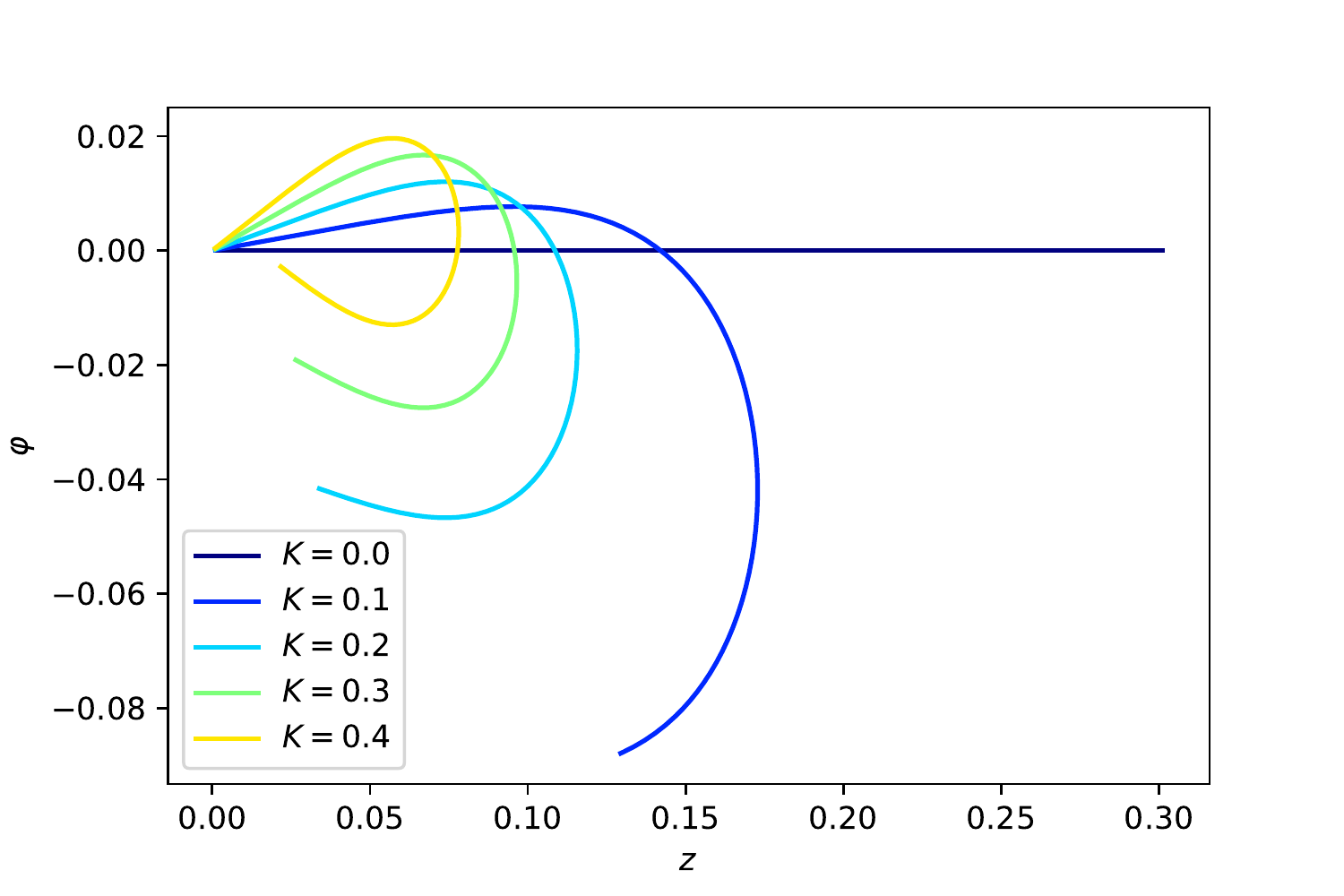}
	\caption{$K$ dependence ($r_m = 10$)}
	\label{fig:iAdSBH_hyp_K1}
	\end{minipage}
\end{figure}

The plots for different masses are shown in Figure \ref{fig:iAdSBH_hyp_M1}.
For larger masses the D5-brane bends tightly.
For $r_\text{m} = 200$ case, the location of the horizon is $z_\text{h}\approx 0.07$ and the D5-brane avoids to exceed this line.
Then we can find that the probe brane does not enter the inner of the black holes.

\begin{figure}[h]
	\begin{minipage}[t]{0.5\linewidth}
	\includegraphics[width=\linewidth]{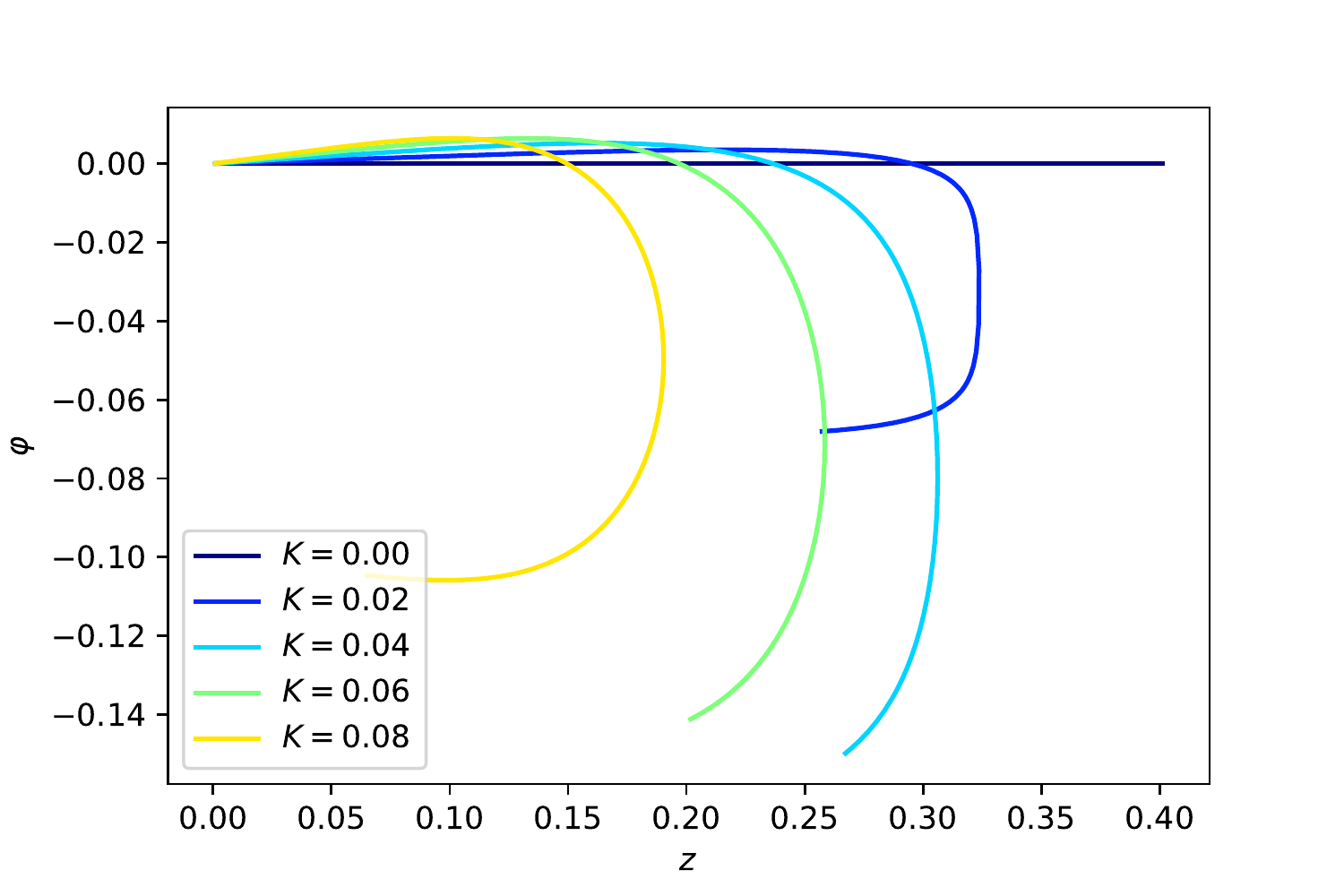}
	\caption{$K$ dependence for small values of $K$ ($r_m = 10$)}
	\label{fig:iAdSBH_hyp_K2}
	\end{minipage}
\hspace{0.01\linewidth}	
	\begin{minipage}[t]{0.5\linewidth}
	\includegraphics[width=\linewidth]{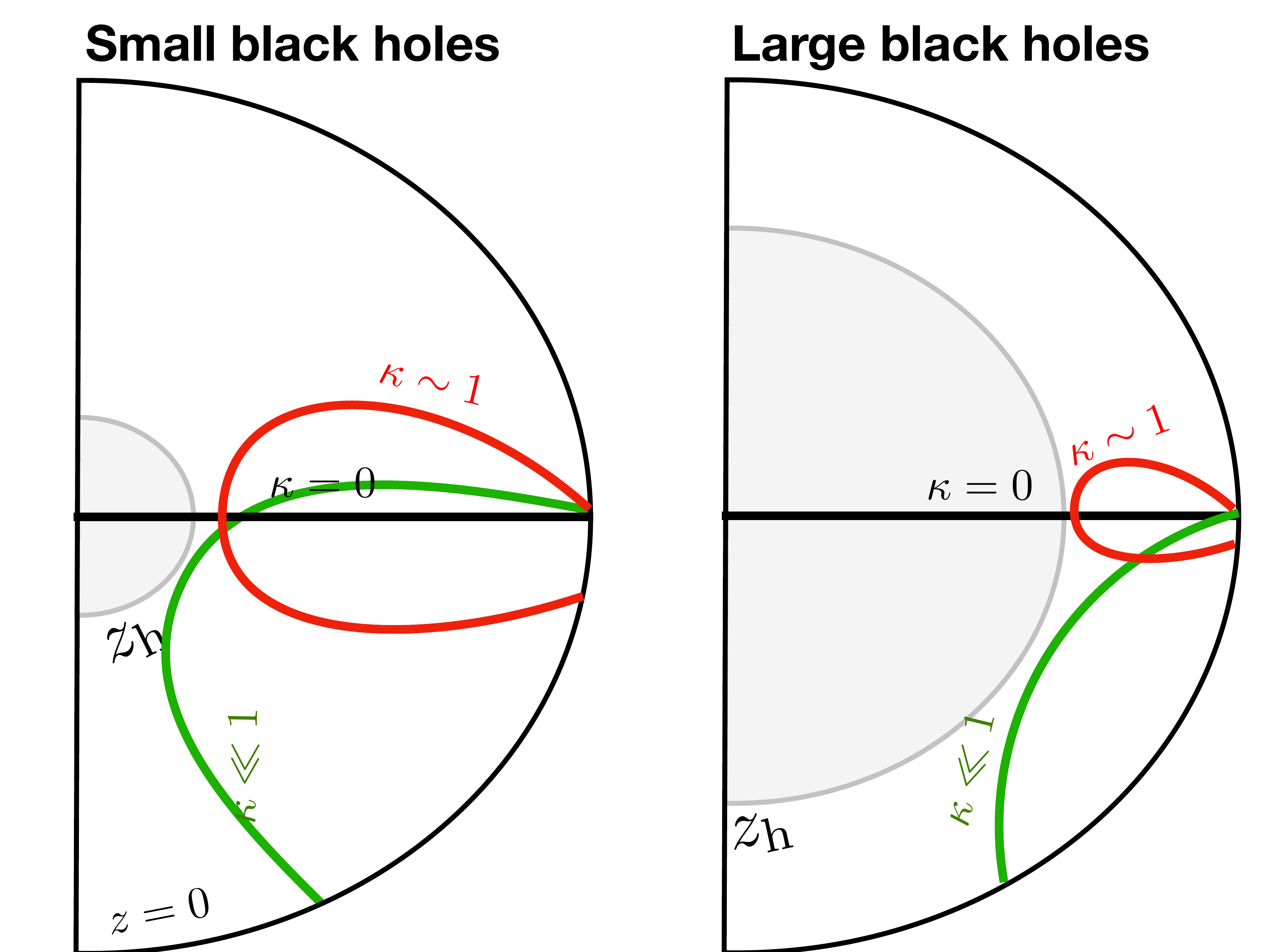}
	\caption{A roughly sketch of the form of the probe D5-brane}
	\label{fig:iAdS_hyp1_sketch}
	\end{minipage}
\end{figure}

The behavior for various values of the flux is drawn in Figure \ref{fig:iAdSBH_hyp_K1} and Figure \ref{fig:iAdSBH_hyp_K2}.
These figures are plotted for mass $r_\text{m}=10$. 
From this results we see that the D5-brane can enter the horizon only in the $K=0$ case. 
For small $K$ values the degrees of slope is small at the boundary ($z=0$) and the probe D5-brane can approach the horizon closer.
As we can see from Figure \ref{fig:iAdSBH_hyp_K2} the D5-brane bend tightly to avoid the horizon for small values of gauge fluxes.
For example the horizon is located at $z_\text{h} = 0.3242$ for mass $r_\text{m} = 10$.

A roughly sketch of the whole form of the D5-brane is shown in Figure \ref{fig:iAdS_hyp1_sketch}.
The circumference of a semi-circle represents the boundary of the AdS spacetime and its interior is the bulk spacetime.
The shaded area represents the inner of the black hole and its border is the black hole horizon.
For larger black holes the horizon becomes larger and the D5-brane does not penetrate the horizon.
Then the D5-brane bends in the vicinity of the boundary.
In the case \cite{Nagasaki:2011ue} there are only one defect on the boundary theory.
However we have two interfaces in the boundary theory.
The gauge group of the boundary gauge theories differ in the both sides of the interface and its difference depends on $k$.
Therefore the boundary space is separated into three regions where three different gauge theories with gauge groups, for example SU($N$), SU($N-k$), and SU($N$), live.

As a consequence, we find that the boundary gauge theory has one interface for $K=0$ and there are two gauge theories.
However, unlike the result of Ref.\cite{Nagasaki:2011ue}, the gauge groups of these two gauge theories are the same because $K=0$ means $k=0$.

For $K\neq0$ there are two interfaces, they separate the boundary into two regions where  gauge theories have different gauge groups by $k$, namely, SU($N$) and SU($N-k$).

\section{Calculation of the DBI action}\label{sec:DBIact}
\begin{figure}[h]
\begin{center}
	\captionsetup{width=15cm}
	\includegraphics[width=14cm]{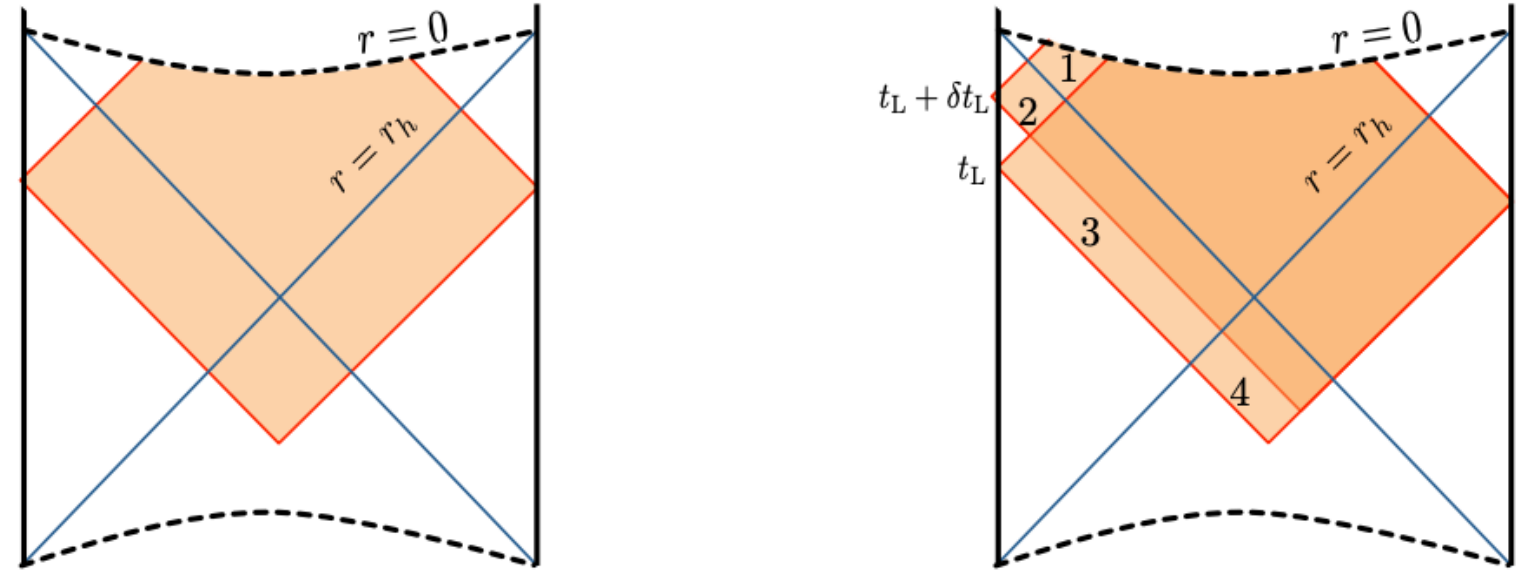}
	\caption{Penrose diagram of the AdS black hole. 
	The blue diagonal lines represent the black hole horizon. 
	(left) The shaded rectangular region represents the Wheeler-de Witt patch (WdW). 
	(right) Two WdW in different left CFT time $t_L$. 
	The regions 1,2,3 and 4 represent the difference of these two patches. 
	The contributions from regions 2 and 3 cancel. 
	The contribution from region 4 becomes smaller as time develops. 
	Then the integration over only the region 1 is relevant for our analysis}
\label{fig:AdSBH_WDW}
\end{center}
\end{figure}

To study complexity according to CA relation, we need the action integrated in the region called WDW patch.
This time development is obtained by integrating inside the horizon (Figure \ref{fig:AdSBH_WDW}).

In the previous section, we found the probe D5-brane can enter the horizon only for the case of no gauge flux.
Then the presence of this probe brane affects only in the zero flux case.

The action for describing this system consists of a bulk term, a surface term and a DBI term. 
In order to find the effect of the probe D5-brane we have to calculate the last term. 
That is shown in eq.\eqref{eq:AdSBH_S_D5}.
The development of this term is 
\begin{equation}\label{eq:IntinWDW}
\frac1{T_5}\frac{dS_\text{D5}}{dt}
=  -(4\pi)^2\sqrt{1+\kappa^2}\int_0^{r_\text{h}} dr (\sin^2\vartheta)r^2
	\Big(\sqrt{1+r^2\vartheta'(r)^2f(r)} - Kr^2\vartheta'(r)\Big).
\end{equation}
\paragraph{$K=0$ case}
When there is no gauge flux, the above integral is simply, 
\begin{equation}
\int_0^{r_\text{h}} drr^2 = \frac13r_\text{h}^3,\;\;
r^2f(r) = r^4 + r^2 - r_\text{m}^2 = 0;\;\;
r_\text{h}^2 = \frac12(-1+\sqrt{1+4r_\text{m}^2}).
\end{equation}
Then the time development of the action in this case is 
\begin{equation}\label{eq:action_interface_K0}
\frac1{T_5}\frac{dS_\text{D5}}{dt}
= -\frac{4\sqrt2\pi^2}3(-1+\sqrt{1+4r_\text{m}^2})^{3/2}.
\end{equation}
This corresponds to the result for no flux ((4.13) of \cite{Abad:2017cgl}).
For large masses, this behaves like $dS_\text{D5}/dt\propto r_\text{m}^{3/2}\propto M^{3/4}$.

We confirmed that our approach gives the same result to the previous works for $k=0$.
On the other hand, for $k\neq0$ case, we know from the result of section \ref{sec:iAdSBH_sol} the prove D5-brane does not exist in the inner of the horizon.
Then we can see in eq.\eqref{eq:IntinWDW} the existence of this probe brane does not affect the growth of complexity because it is integrated inside of the horizon ($0\leq r\leq r_\text{h}$).

\section{Fundamental string}\label{sec:iAdSBH_fstring}
We consider a fundamental string whose one side sits on the interface and the other side extends to the AdS boundary with the latitude angle $\varphi_0$.
We parametrize the string worldsheet by coordinates $(\tau,\sigma)$.
We consider a static string then the embedding is expressed by
\begin{align}
X^0(\tau,\sigma) &= t(\tau),\nonumber\\
X^1(\tau,\sigma) &= r(\sigma) = 1/z(\sigma),\nonumber\\
X^2(\tau,\sigma) &= \vartheta(\sigma) = \frac\pi2-\varphi(\sigma),\nonumber\\
X^\mu(\tau,\sigma) &= 0\;(3\leq \mu \leq9),\qquad
\sigma\in(0,\sigma_1).
\end{align}
Denoting the form of the D5-brane as $\varphi = S(z)$, the boundary conditions at the both sides are
\begin{subequations}\label{eq:bc_string_interfaceBH}
\begin{align}
&\sigma = 0:
&& 
\varphi(0) = \varphi_0,\;\;
\varphi'(\sigma)\big|_{\sigma=0} = 0,\\
&\sigma = \sigma_1:
&&
\varphi(\sigma_1) = S(z(\sigma_1)),\;\;
S'(z)\varphi'(\sigma)\big|_{\sigma=\sigma_1} 
	+ z'(\sigma)\big|_{\sigma=\sigma_1} = 0.
\end{align}
\end{subequations} 
These mean we imposed the Neumann boundary condition along the boundary of AdS$_5$ and the Dirichlet boundary condition for its transverse direction.
\footnote{
These boundary conditions are obtained as follows.
First the Dirichlet boundary conditions at both sides are
\begin{equation*}
\varphi(0) = \varphi_0,\;\;
\varphi(\sigma_1) = \varphi(z(\sigma_1)) =: \varphi(z_1).
\end{equation*}
The remaining conditions, the Neumann boundary conditions, are the condition for the string to be perpendicular to the boundary.
Then, at $\sigma = 0$, 
\begin{equation*}
\frac{d\varphi}{d\sigma}\Big|_{\sigma=0} = 0.
\end{equation*}
At the other end, the string is perpendicular to the D5-brane whose embedding is $\varphi = S(z)$.
Therefore, the condition under which the D5-brane and the string meet perpendicularly gives
\begin{equation*}
\frac{dS}{dz}\frac{d\varphi(\sigma)/d\sigma}{dz(\sigma)/d\sigma} = -1,\;\;
\frac{dS}{dz}\frac{d\varphi(\sigma)}{d\sigma} 
+ \frac{dz(\sigma)}{d\sigma} = 0.
\end{equation*}
}

The Polyakov action is, by denoting the derivative by 
$t$ as $\dot{\;\;} := \partial/\partial t$ and 
$\sigma$ as ${}^\prime := \partial/\partial\sigma$,
\begin{align}
S_\text{P}
&= \frac1{4\pi\alpha'}\int d\tau d\sigma 
\sqrt{h}h^{ab}\partial_a X^\mu \partial_b X^\nu g_{\mu\nu}\nonumber\\
&= \frac1{4\pi\alpha'}\int d\tau d\sigma g_{\mu\nu}
(\dot X^\mu \dot X^\nu + X'^\mu X'^\nu)\nonumber\\
&= \frac1{4\pi\alpha'}\int d\tau d\sigma\Big[
f(r)\dot t^2 + \frac{r'^2}{f(r)} 
  + r^2\vartheta'^2\Big],
\end{align}
where in the second equality we chose the conformal gauge $h = \text{diag}[1,1]$ and the third equality we used the Euclidean version of the AdS black hole metric \eqref{eq:dsAdSBH}:
\begin{align}
ds_\text{AdSBH}^2
&= f(r)dt^2 + \frac{dr^2}{f(r)} + r^2d\Omega_3^2 + d\Omega_5^2,\nonumber\\
& f(r) = 1 - \frac{r_\text{m}^2}{r^2} + r^2,\qquad
r_\text{m}^2 = \frac{8M}{3\pi},\qquad
d\Omega_3^2 = d\vartheta^2 
  + \sin^2\vartheta(d\psi_1^2 + \sin^2\psi_1d\psi_2^2).
\end{align}
The equations of motion for variables, $t, \vartheta$ and $r$, are
\begin{align}
&\ddot t = 0,\;\; t = \tau;\;\;
0 = \frac{d}{d\sigma}(\partial_{\vartheta'}\mathcal L) 
 - \partial_\vartheta\mathcal L
= \frac{d}{d\sigma}(2r^2\vartheta'),\;\;
\vartheta' = \frac{L}{r^2};\nonumber\\
&0 = \frac{d}{d\sigma}\Big(\frac{2r'}{f(r)}\Big)
  - \Big(f'(r)\dot t^2 - \frac{r'^2f'(r)}{f(r)^2} 
  + 2r\vartheta'^2\Big)
= \frac{2r''}{f(r)} - f'(r) - \frac{r'^2f'(r)}{f(r)^2}
  - \frac{2L^2}{r^3}.
\end{align}
The last equation is by multiplying $r'$,
\begin{equation}\label{eq:Virasoro_const}
\frac{(r'^2)'}{f(r)} - r'f'(r) + r'^2\frac{d}{d\sigma}\frac1{f(r)} - \frac{2L^2r'}{r^3} 
= \frac{d}{d\sigma}\Big(\frac{r'^2}{f(r)} - f(r) + \frac{L^2}{r^2}
\Big),\;\;
\frac{r'^2}{f(r)} - f(r) + \frac{L^2}{r^2}
= \text{const.}
\end{equation}
This constant is determined by the following.

\paragraph{Virasoro constraint}
The energy momentum tensor is 
\begin{equation}
T_{\alpha\beta} 
= \partial_\alpha X\cdot\partial_\beta X - \frac12h_{\alpha\beta}h^{\gamma\delta}\partial_\gamma X\cdot\partial_\delta X = 0.
\end{equation}
Here we chose the conformal gauge then $h_{00} = h_{11} = 1$.
Each component gives the Virasoro constraint:
\begin{align*}
0 = T_{00} 
&= \partial_0X\cdot\partial_0X 
  - \frac12(\partial_0X\cdot\partial_0X + \partial_1X\cdot\partial_1X)
= \frac12(\partial_0X\cdot\partial_0X - \partial_1X\cdot\partial_1X),\\
0 = T_{11} 
&= \partial_1X\cdot\partial_1X 
  - \frac12(\partial_0X\cdot\partial_0X + \partial_1X\cdot\partial_1X)
= - T_{00}.
\end{align*}
Then the Virasoro constraint for our problem is 
\begin{equation}
0 = \partial_0X\cdot\partial_0X - \partial_1X\cdot\partial_1X 
= f(r)\dot t^2 - \frac{r'^2}{f(r)} - r^2\vartheta'^2
= f(r) - \frac{r'^2}{f(r)} - \frac{L^2}{r^2}.
\end{equation}
Then the constant in equation \eqref{eq:Virasoro_const} is determined.
By solving it for $r'$, we obtain
\begin{equation}
r' = -\sqrt{f(r)\Big(f(r) - \frac{L^2}{r^2}\Big)},
\end{equation}
where we choose the minus sign since our parametrization.
By the Virasoro constraint obtained above, the Polyakov action is rewritten as 
\begin{equation}
\frac{dS_\text{P}}{dt} 
= \frac1{4\pi\alpha'}\int_{r_1}^{\infty} d\sigma\Big[
	f(r) + \frac{r'^2}{f(r)} + r^2\vartheta'^2\Big]
= \frac1{2\pi\alpha'}\int_{r_1}^{\infty} d\sigma f(r).
\end{equation}

\begin{figure}[h]
	\begin{minipage}[h]{0.4\linewidth}
	\includegraphics[width=\linewidth]{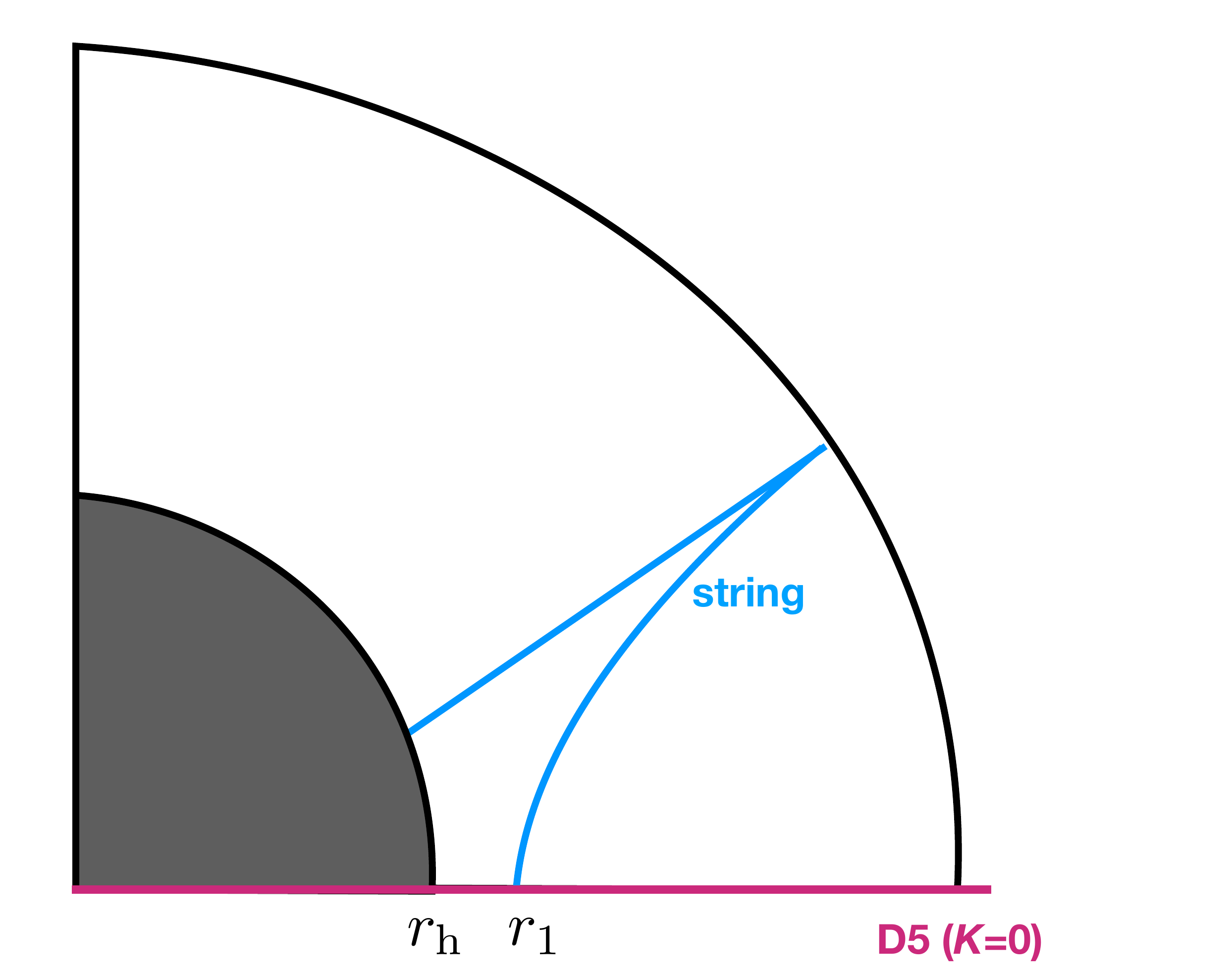}
	\caption{A fundamental string ending on the D5-brane}
	\label{fig:stringonD5}
	\end{minipage}
\hspace{0.01\linewidth}
	\begin{minipage}[h]{0.5\linewidth}
	\includegraphics[width=\linewidth]{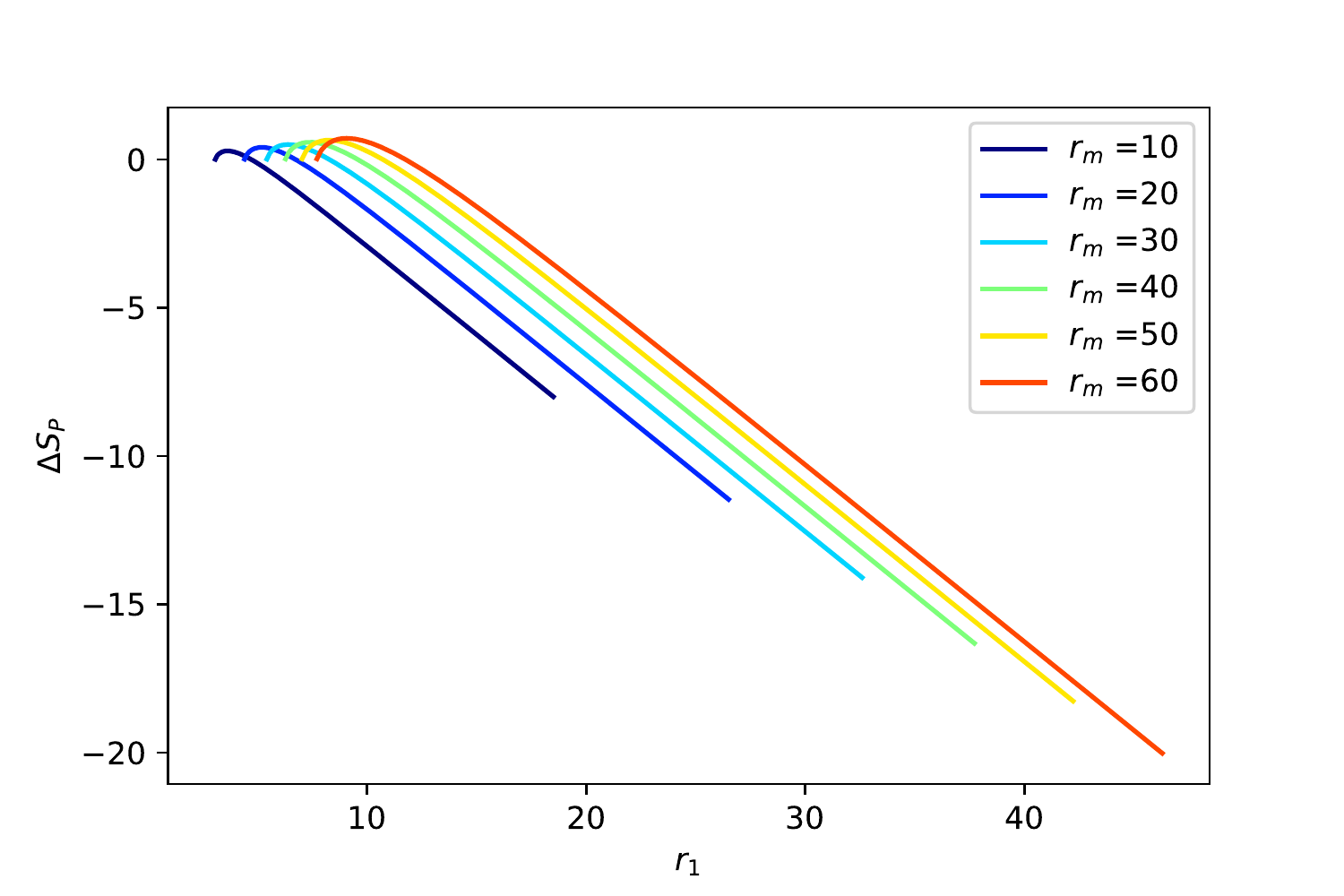}
	\caption{Polyakov action as a function of $r_1$}
	\label{fig:S_intAdSBHp_r0}
	\end{minipage}
\hspace{0.01\linewidth}
	\begin{minipage}[h]{0.5\linewidth}
	\includegraphics[width=\linewidth]{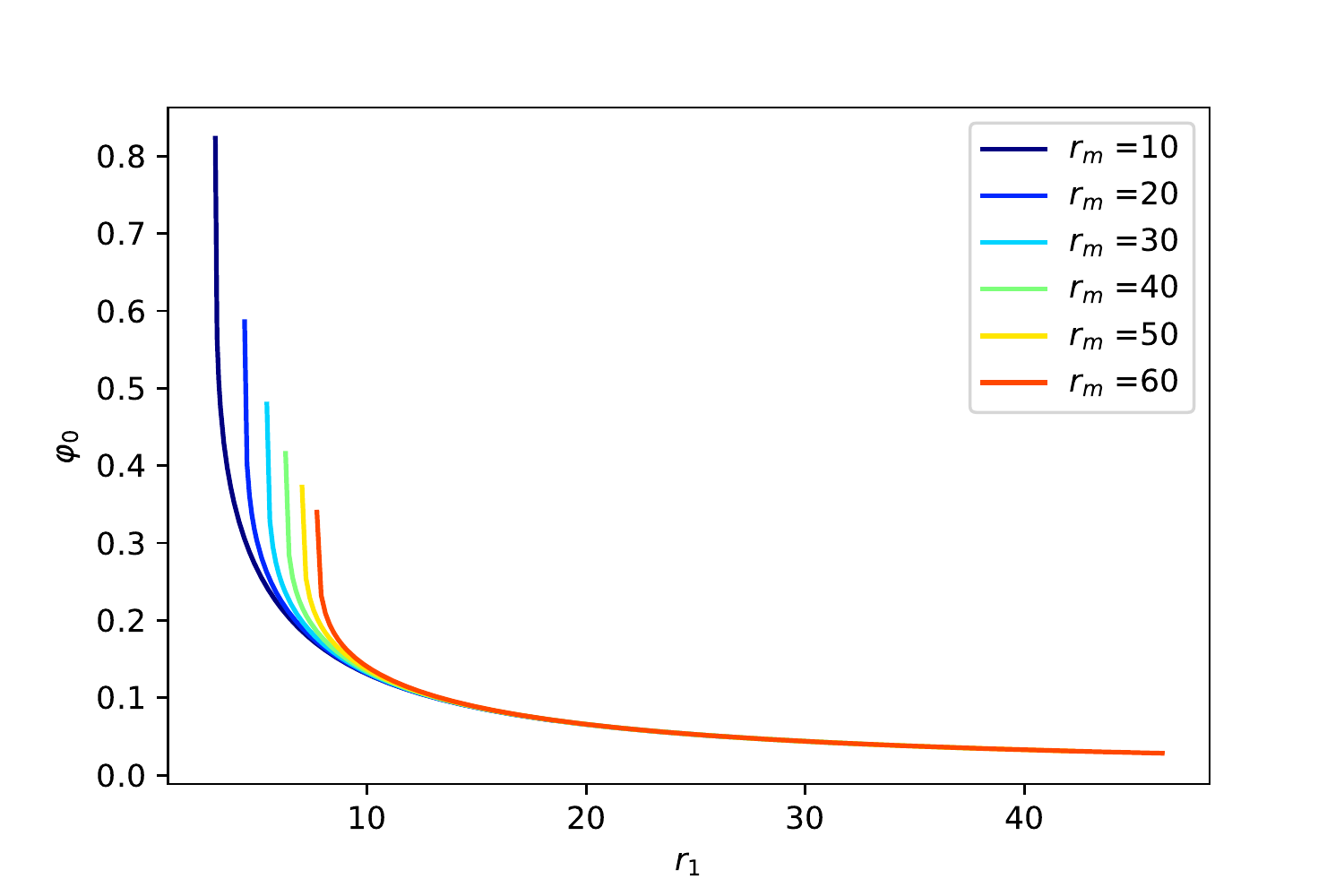}
	\caption{$r_1$ dependence of $\varphi_0$}
	\label{fig:phi0_r1}
	\end{minipage}
\hspace{0.01\linewidth}
	\begin{minipage}[h]{0.5\linewidth}
	\includegraphics[width=\linewidth]{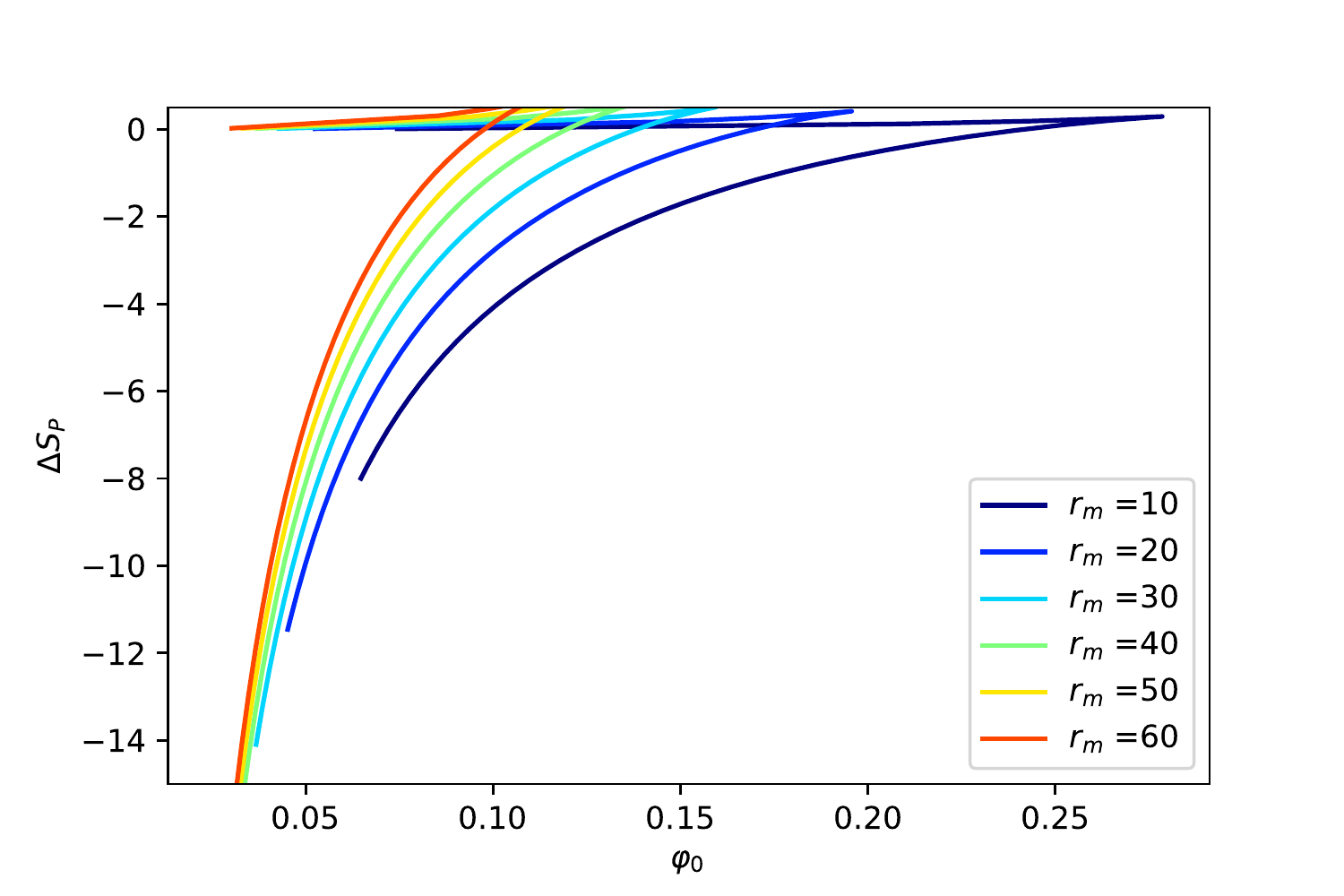}
	\caption{Polyakov action as a function of $\varphi_0$}
	\label{fig:S_intAdSBHp_phi0}
	\end{minipage}
\end{figure}

\paragraph{$K = 0$ case}
For simplicity we consider the case where there is no flux.
We would like to consider the situation where the string ends on the D5-brane outside the horizon.
The abstract form of the string is depicted in Figure \ref{fig:stringonD5}.
In this case $\varphi(z) = S(z) = 0$, then $S'(z) = 0$.
The boundary condition \eqref{eq:bc_string_interfaceBH} is simplified as
\begin{equation}
\varphi(0) = \varphi_0,\;\;
\varphi(\sigma_1) = 0,\;\;
\frac{d\varphi}{d\sigma}\Big|_{\sigma=0} = 0,\;\;
\frac{dz}{d\sigma}\Big|_{\sigma=\sigma_1} = 0.
\end{equation}
From this boundary condition
\begin{equation}
\frac{dr}{d\sigma}\Big|_{r=r_1} 
= -\frac{\sqrt{f(r)(r^2f(r)-L^2)}}{r} = 0,\;\;
\therefore L^2 = r_1^2f(r_1).
\end{equation}
Since the string now exists outside the horizon $f(r)\geq0$, then
\begin{equation}\label{eq:Pact_string_interBH1}
S_\text{P} 
= \frac{T}{2\pi\alpha'}\int_{r_1}^{\infty} dr\sqrt\frac{r^2f(r)}{r^2f(r) - r_1^2f(r_1)}.
\end{equation}
We also define the following action in the case where there is no probe D5-brane and the string directly enters the horizon (Figure \ref{fig:stringonD5}).
\begin{equation}\label{eq:Pact_string_interBH2}
S_\text{P0} 
= \frac{T}{2\pi\alpha'}\int_{r_\text{h}}^{\infty} dr,\;
(\because L = 0),\;
r_\text{h} = \frac{-1+\sqrt{4r_\text{m}^2+1}}{2} < r_1.
\end{equation}
Since the integral \eqref{eq:Pact_string_interBH1} diverges, we define the effect of the probe D5-brane by subtracting the action \eqref{eq:Pact_string_interBH2},
\begin{align}
\frac{2\pi\alpha'}{T}\Delta S_\text{P} 
= \int_{r_1}^{\infty} dr\sqrt\frac{r^2f(r)}{r^2f(r)-r_1^2f(r_1)}
- \int_{r_\text{h}}^{\infty} dr
= \int_{r_1}^{\infty} dr\Bigg(\sqrt\frac{r^2f(r)}{r^2f(r)-r_1^2f(r_1)} - 1\Bigg)
- r_1 + r_\text{h}.
\end{align}
By the integration, $\varphi_0$ is obtained as a function of $r_1$ as (Figure \ref{fig:phi0_r1})
\begin{align}
\frac{d\vartheta}{dr}
&= \frac{-r}{\sqrt{r^2f(r)(r^2f(r) - L^2)}},\;\;
\vartheta
= \frac\pi2 - \int_{r_1}^\infty\frac{rdr}{\sqrt{f(r)(r^2f(r) - L^2)}},\nonumber\\
\varphi_0
&= \int_{r_1}^\infty\frac{rdr}{\sqrt{(1+r^2-r_\text{m}^2/r^2)(1+r^2+r_1^2)(r^2-r_1^2)}}.
\end{align}
The result of the numerical calculation is shown in Figure \ref{fig:S_intAdSBHp_phi0}.

This behavior of the potential function for different positions of string's edge is shown in Figure \ref{fig:S_intAdSBHp_phi0}.
The horizontal axis is the location of the string's edge at the boundary.
$\phi_0$ represents the location of the interface.
In this figure the negative values region is only valid. 
If the value of the action $\Delta S_p$ exceeds zero, a phase transition occurs and the string directly goes into the black hole (Figure \ref{fig:stringonD5}). 
We can see that for larger black holes the transition point becomes small and the transition becomes likely to occur.

\section{Summary and discussion}\label{Sec:Discussion}
In this paper we treated the interface which is a co-dimension one object realized by the probe D5-brane in AdS black hole spacetime.
We obtained the interface solution in the AdS black hole spacetime by the numerical calculation.
The boundary condition at infinity is given in the same way for the AdS$_5\times S^5$ case imposed in \cite{Nagasaki:2011ue}, that is, the effect is the black hole is neglected and approximated by the flat case solution.
Approaching the black hole, the D5-brane is bent.
The behavior of the D5-brane changes according the strength of the flux in the same as the flat AdS case.
The whole behavior can be seen in Figure \ref{fig:iAdS_hyp1_sketch}.
A new result of this research is the transparency of the black hole horizon, that is, the probe brane enter the horizon only the case where the brane has no gauge flux and if the brane has non-zero gauge flux then it bends tightly to avoid the horizon.

For the case which has no flux we also calculated the quark-interface potential.
This is a correlation function between a local object (a quark) and a nonlocal object (interface or defect).
This is given by the action of the fundamental string connecting the D5-brane and the boundary.
Since this action diverges, we defined the effect of the interface by subtracting the action which does not have the interface.
We obtained the effect of the interface as a function of the position of the string endpoint which represent the particle at infinity.
It has phase transition because of the selection of the smaller energy.
This behavior is shown in Figure \ref{fig:S_intAdSBHp_phi0}.
The horizontal line represents the distance from the interface.
In this figure effective values are on the negative region.
If the value of the action $\Delta S_p$ exceeds zero, a phase transition occurs and the string directly goes into the black hole (Figure \ref{fig:stringonD5}).	
For larger black holes, the transition point of $\phi_0$ becomes small.

For a future work we are interested in a moving defect \cite{Janiszewski:2011ue} or not stationary case.
For example, the black hole can have the angular momentum or the interface may be time developing.
In these cases there is a possibility that the probe brane enter the horizon and have an effect to the black hole interior.
Then in this case this is related to the problem of complexity because its time development is calculated in the Wheeler-DeWitt (WDW) patch including the inner of the horizon.

\section*{Acknowledgement}
I would like to thank Satoshi Yamaguchi, UESTC members and people discussing at  International Workshop on "Theoretical Particle Physics 2018" for useful comment.

\providecommand{\href}[2]{#2}\begingroup\raggedright\endgroup

\end{document}